\documentclass{aa}  
\usepackage{graphicx}
\usepackage{txfonts}
\usepackage{lipsum}
\usepackage{subcaption}         
\usepackage{lscape}            
\usepackage{placeins}           
\usepackage{xcolor}
\usepackage{hyperref}
\usepackage{comment}
\usepackage{stfloats}
\usepackage{lineno}

\begin{document} 

   \title{The interstellar flux gap: From dust to kilometer-scale objects}
   \titlerunning{Interstellar flux gap}
   
   \author{E. Peña-Asensio \inst{1,2}
          \and
          D. Z. Seligman\inst{3}
          }
    \institute{
        Department of Applied Mathematics and Aerospace Engineering, Universitat d'Alacant, 03690 Alacant, Spain
        \email{eloypa@ua.es}
        \and
        Department of Aerospace Science and Technology, Politecnico di Milano, Via La Masa 34, 20156 Milano, Italy\\
        \and
        Department of Physics and Astronomy, Michigan State University, East Lansing, MI 48824, USA
        \\
    }
 
    \abstract
    {Three kilometer-sized interstellar objects (ISOs) have been detected transiting the Solar System, and spacecraft have directly measured micrometer-scale interstellar dust (ISD). Yet no intermediate-size interstellar meteoroids have been identified in current meteor surveys.}
    {We test whether a power-law flux extrapolation connecting spacecraft ISD and kilometer-scale ISOs is consistent with meteor surveys, and we quantify the expected interstellar impacting flux based on various observational reports.}
    {We compiled differential fluxes and limits from spacecraft ISD, radar and optical meteor surveys, and theoretical estimates. We  evaluated the power-law size–frequency fits, computed the 3I-like flux, and compared measured fluxes to predictions.}
    {The spacecraft-measured dust flux exceeds extrapolations constrained by meteor surveys and kilometer-scale ISOs by $\sim$2--7 orders of magnitude. An $r^{-3.0}$ fit combining spacecraft ISD detections with kilometer-scale ISOs overpredicts the number of meteors with hyperbolic orbits, whereas slopes of $r^{-2.7}$--$r^{-2.3}$ (derived from radar and optical meteor upper limits, respectively) instead yield interplanetary-to-interstellar flux ratios of $10^{3}$--$10^{6}$.}
    {A simple power-law from ISD to ISOs is inconsistent with meteor survey constraints and yields unrealistic predictions for interstellar meteoroids. The data reveal a gap between submicron dust entrained in the Local Interstellar Cloud (LIC) and macroscopic bodies ejected from planetary systems. This gap may reflect distinct origins and destruction-transport processes rather than a continuous size-frequency distribution. This would imply either the dominance of a small-particle LIC component or the need to reassess spacecraft dust fluxes.}


   \maketitle

\nolinenumbers

\section{Introduction}
There have been three detections of macroscopic kilometer-scale interstellar objects (ISOs) traversing the Solar System: 1I/‘Oumuamua \citep{Williams17}, 2I/Borisov \citep{borisov_2I_cbet}, and 3I/ATLAS \citep{Denneau2025}. There have also been in situ measurements of much smaller --- approximately micron-scale --- interstellar dust (ISD) particles by spacecraft. In particular, the Ulysses and Galileo spacecraft measurements have provided a calibration of the ISD flux \citep{Grun1993, Grun1997, Grun2000, Landgraf2000}. Attempts have been made to fit a power-law extrapolation for the size frequency distribution of ISD to the meter scale in order to estimate upper limits of interstellar meteoroids \citep{Musci12}. It is natural to extend this size-frequency distribution to that of ISOs implied by 1I/‘Oumuamua into a single power-law spanning approximately ten orders of magnitude, for example as performed in Figure~19 of \citet{Jewitt2023ARAA}. 

Extrapolating this slope predicts the existence of numerous millimeter- to meter-scale interstellar impactors that should already have been detected by current meteor networks. However, the results from extant meteor surveys do not support this prediction. The Canadian Meteor Orbit Radar (CMOR) contains only five $3\sigma$ hyperbolic candidates among 11 million orbits \citep{Froncisz2020PSS19004980F}. The Global Meteor Network (GMN) contains no secure millimeter-sized interstellar meteoroids and sets an upper limit ratio of $\sim10^{-6}$ relative to the number of interplanetary meteors \citep{Wiegert2025}.

One of the most reputed and extensive fireball compilations -- the European Fireball Network (EN) catalog \citep{Borovicka2022AA667A158B} -- reports 15 ($\sim$2\%) hyperbolic events at a minimum 1$\sigma$ confidence level, though this includes two events at a 3$\sigma$ confidence. Nevertheless, the authors do not consider the events truly interstellar. Meteor and fireball databases typically list a small number of formally hyperbolic events (1-12$\%$), but they are attributed to measurement error rather than being of a true interstellar origin \citep{Hajdukova2020b, Hajdukova2024}. In Table~\ref{tab:interstellar} we show a compilation of interstellar candidate impactors and confirmed macroscopic interstellar visitors. As expected, orbital solutions for fireballs are more accurate than those for radar meteors, but their uncertainties remain well above the precision achieved for telescopic observations of kilometer-sized ISOs. Candidate meteoroids with hyperbolic orbits detected by radar are exclusively prograde (15°--67°), whereas the optical fireball candidates are retrograde, except for one case spanning ~60°--170°. In contrast, the confirmed ISOs display no such preference: 1I/‘Oumuamua (122.7°; \citealp{JPLHorizonsOumuamua}), 2I/Borisov (44.1°; \citealp{JPLHorizonsBorisov}), and 3I/ATLAS (175.1°; \citealp{JPLHorizonsATLAS}). They cover both prograde and retrograde regimes and are consistent with an isotropic distribution around 90°. However, radar favors lower-speed prograde orbits, while optical surveys favor luminous high-speed retrograde entries, so the inclination contrast may reflect detection bias and should not be overinterpreted.

\begin{figure}
    \includegraphics[width=1.\linewidth, trim=0 10 35 35, clip]{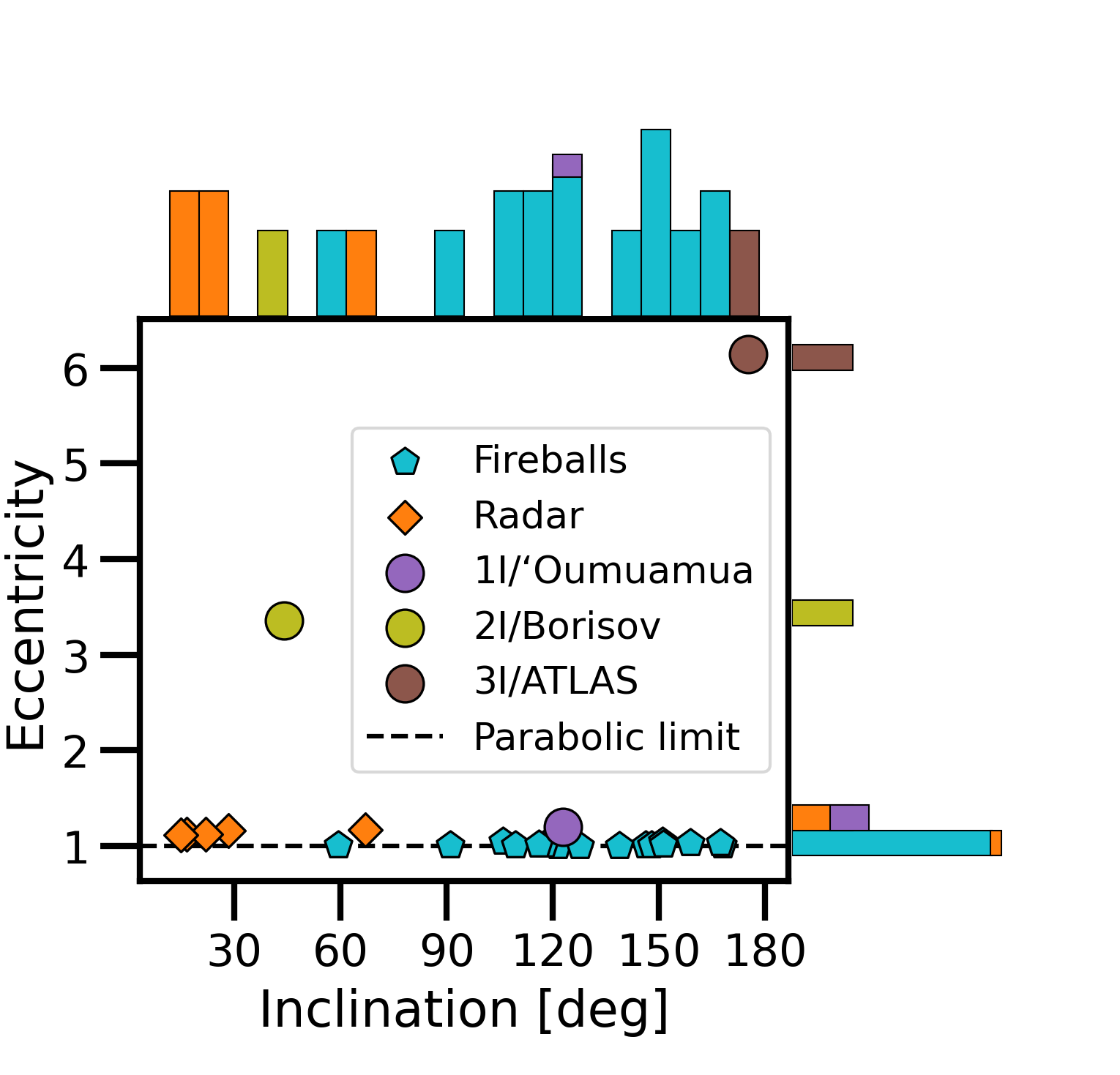}
    \caption{Eccentricity versus inclination for hyperbolic fireballs reported in the EN catalog \citep{Borovicka2022AA667A158B}, hyperbolic meteor detections from CMOR radar surveys \citep{Froncisz2020PSS19004980F}, and the three known ISOs 1I/‘Oumuamua, 2I/Borisov, and 3I/ATLAS \citep{JPLHorizonsOumuamua,JPLHorizonsBorisov,JPLHorizonsATLAS}. Marginal histograms display the distribution of inclination (top) and eccentricity (right) for each dataset in logarithmic scale using colors consistent with the scatter plot. The dashed horizontal line marks $e=1$.
    }
    \label{fig:hyp_ecc_inc}
\end{figure}

A similar contrast is seen in eccentricity. The meteors cluster just above the parabolic limit ($e \leq 1.17$), while the ISOs show far higher hyperbolic excess velocities and eccentricity values (e.g., $e=1.20$ for 1I, $e=3.36$ for 2I, $e=6.14$ for 3I). Figure~\ref{fig:hyp_ecc_inc} shows eccentricity versus inclination for meteoroids with hyperbolic orbits and the three known ISOs. 

The perceived tension -- namely, the absence of an observed intermediate-sized population -- arises from a potentially invalid assumption that the spacecraft ISD sample and macroscopic ISOs belong to a single and uniform population. The spacecraft data consist of submicron grains entrained in the Local Interstellar Cloud (LIC) flow that are strongly coupled to interstellar and heliospheric magnetic fields. Their inferred flux depends strongly on selection and mass calibration. Alternative treatments shift the flux by factors on the order of two and contaminate the high-mass tail with interplanetary particles \citep{Baalmann2025AA}. Moreover, no grains larger than $\sim10~\mu$m have been robustly detected in two decades. Destructive processes in the interstellar medium (ISM; sputtering, shattering) may truncate the ISD size distribution on timescales $\lesssim0.6$ Gyr \citep{Jones1994ApJ, Zhukovska2008AA479453Z, Bocchio2014AA570A32B, Zhukovska2016ApJ, Hu2019MNRAS}. This is significantly shorter than the $\sim2.5$ Gyr residence time \citep{Zhukovska2008AA479453Z}, and therefore these processes may deplete large grains before arrival.

Combined with heliopause filtering, radiation-pressure cones, and solar cycle–driven Lorentz modulation, these destructive processes suggest a systematic underrepresentation of submicrometer to approximately millimeter ISD in spacecraft measurements. At the heliospheric boundary, nanometer to submicron grains are excluded by strong electromagnetic coupling. Aggregates in this region are more efficiently removed due to enhanced charging \citep{Sterken2012AA538A102S, Sterken2019SSRv21543S, Sterken2022SSRv21871S}. Solar radiation pressure removes grains for which the ratio of radiation pressure to solar gravity exceeds unity. This creates paraboloid-shaped exclusion zones. At the same time, sub-$0.4~\mu$m particles are strongly deflected by Lorentz forces, which produce a 22-year modulation between focusing and defocusing phases \citep{Sterken2012AA538A102S, Sterken2013AA552A130S}. Both the ISM destruction and heliospheric modulation in concert reshape the incoming ISD population such that its size distribution at 1 AU may no longer reflect that of the LIC, with an expected flattening of the size-frequency distribution on smaller particles.

In contrast, ISOs are macroscopic planetesimals ejected from planetary systems that are effectively decoupled from magnetic forces and travel on gravitationally dominated hyperbolic trajectories. Their local number density and relative velocities and incoming directions should reflect stellar system dynamics \citep{Taylor2025ApJ990L14T,Seligman2018,MoroMartin2022_seeds,Dorsey2025,Marceta2023a,Marceta2023b,Hopkins2023,Hopkins2025}, not the diffuse ISM. However, \citet{Forbes2024} have examined stellar stream–like modulations in ISO kinematics, but these are a second-order effect relative to dynamical heating.

\section{Flux comparison}
We assumed a radius range of 0.22--2.8 km \citep{Jewitt2025ApJ990L2J} and an asymptotic heliocentric speed of 58 km s$^{-1}$ \citep{Seligman2025} for 3I/ATLAS, and we adopted a number density of $10^{-4}\,\mathrm{AU}^{-3}$ for such objects \citep{Taylor2025ApJ990L14T}. As a result, we obtained a flux of
$\sim1.7\cdot 10^{-33}\,\mathrm{m^{-2}\,s^{-1}}$. 

\begin{figure*}
    \centering
    \includegraphics[width=0.8\linewidth, trim=5 20 0 0, clip]{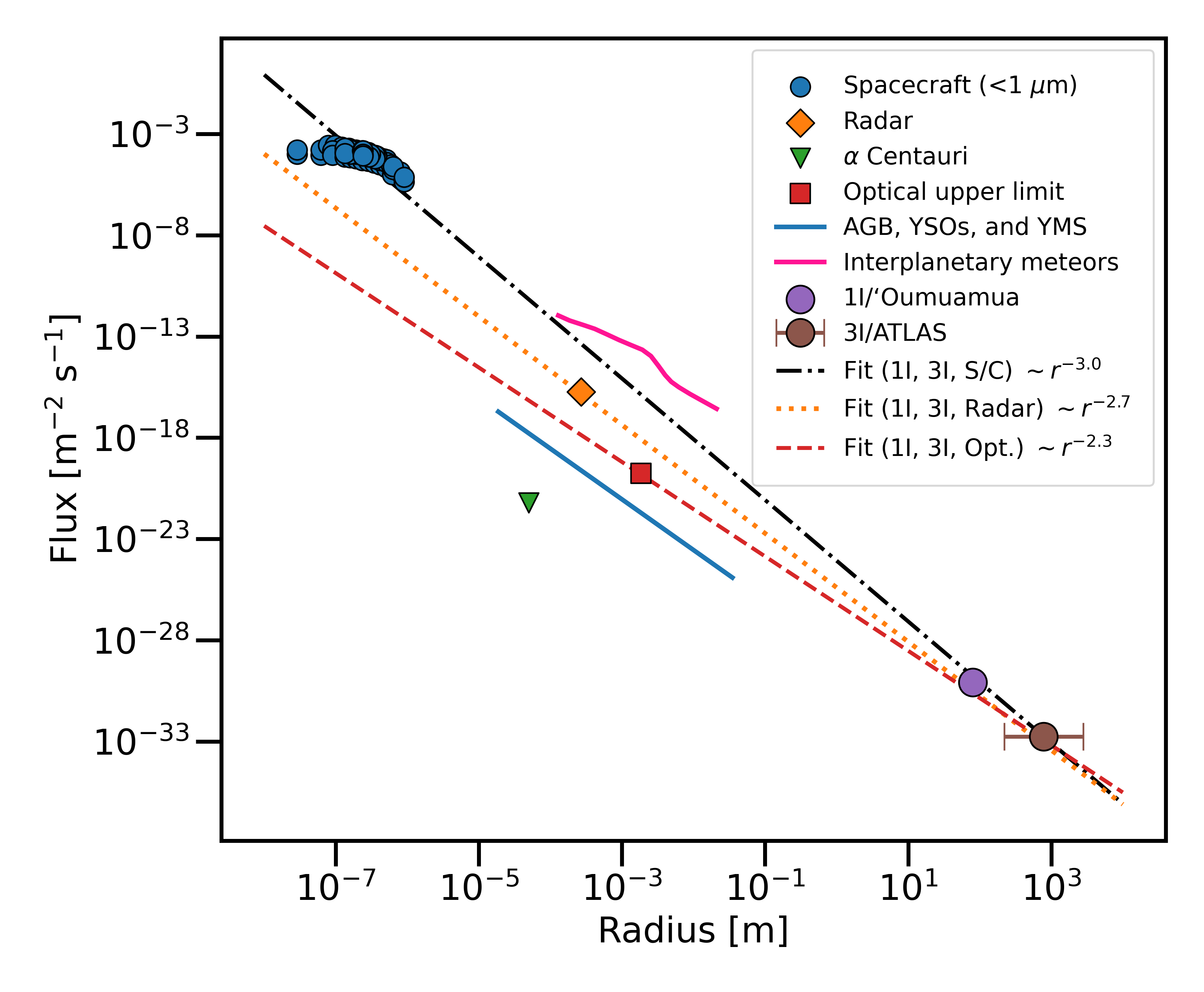}
    \caption{Differential flux versus size based on Ulysses and Galileo spacecraft confident ISD detections \citep{Baalmann2025AA}, hyperbolic meteor radar 3$\sigma$ constraint in CMOR \citep{Froncisz2020PSS19004980F}, estimated $>$100 $\mu$m objects impacting the Earth from $\alpha$ Centauri \citep{GreggWiegert2025}, optical meteor upper limit in GMN \citep{Wiegert2025}, estimated flux at Earth of objects from asymptotic giant branch (AGB) stars, young stellar objects (YSOs), and young main-sequence stars \citep[YMSs;][]{Murray2004}, observed sporadic meteors on Earth with interplanetary orbits \citep{Ceplecha1988BAICz39221C}, 1I/‘Oumuamua-like object flux \citep{Hajdukova2019}, and 3I/ATLAS‐like object (this work). Linear fit fluxes are depicted for $r^{-3.0}$ \citep{Jewitt2023ARAA}, $r^{-2.7}$ to match 1I and 3I with hyperbolic radar meteors, and $r^{-2.3}$ to match 1I and 3I with upper limit hyperbolic optical meteors.
    }
    \label{fig:fluxes}
\end{figure*}

Different object fluxes are depicted in Figure~\ref{fig:fluxes} as a function of size. An immediate takeaway from the flux comparison presented in this figure is that a simple power-law extrapolation from spacecraft ISD to macroscopic ISOs does not hold. If the $r^{-3.0}$ slope proposed by \citet{Jewitt2023ARAA} were valid across all scales, $\sim$0.1 mm interstellar impactors would be as common as our native interplanetary meteors \citep{Ceplecha1988BAICz39221C}, which clearly does not occur.

\begin{figure}
    \includegraphics[width=1.\linewidth]{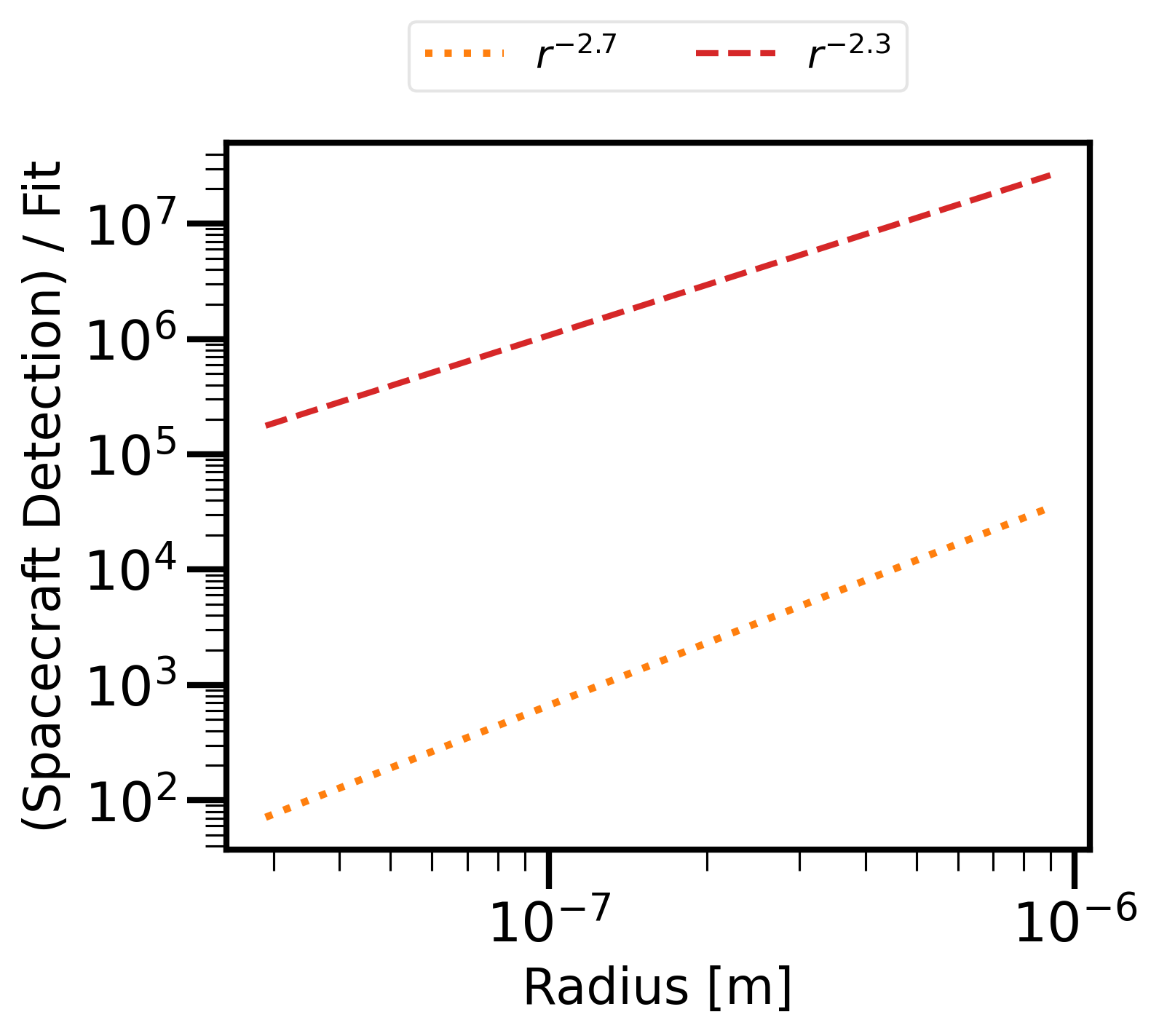}
    \caption{Ratio of the measured differential flux of ISD detections by spacecraft to the corresponding derived power-law fit for interstellar impactors. Linear fit fluxes are depicted for $r^{-2.7}$ to match 1I and 3I with hyperbolic radar meteors and for $r^{-2.3}$ to match 1I and 3I with upper limit hyperbolic optical meteors (see Figure~\ref{fig:fluxes}).
    }
    \label{fig:ratio_spacecraft}
\end{figure}

As depicted in Figure~\ref{fig:ratio_spacecraft}, the power‐law fit matching kilometer‐scale visitors and constrained by hyperbolic radar measurements and kilometer-sized ISOs (\(r^{-2.7}\)) indicates that the ISD flux detected by spacecraft (coming from the LIC) is on average two to four orders of magnitude greater than the extrapolated dust flux coming directly from other planetary systems. In contrast, relative to the fit constrained by the optical meteor upper limit and kilometer-sized ISOs \(r^{-2.3}\), the LIC flux would be approximately five to seven orders of magnitude smaller than the dust flux detected by spacecraft.

\begin{figure}
    \includegraphics[width=1.\linewidth]{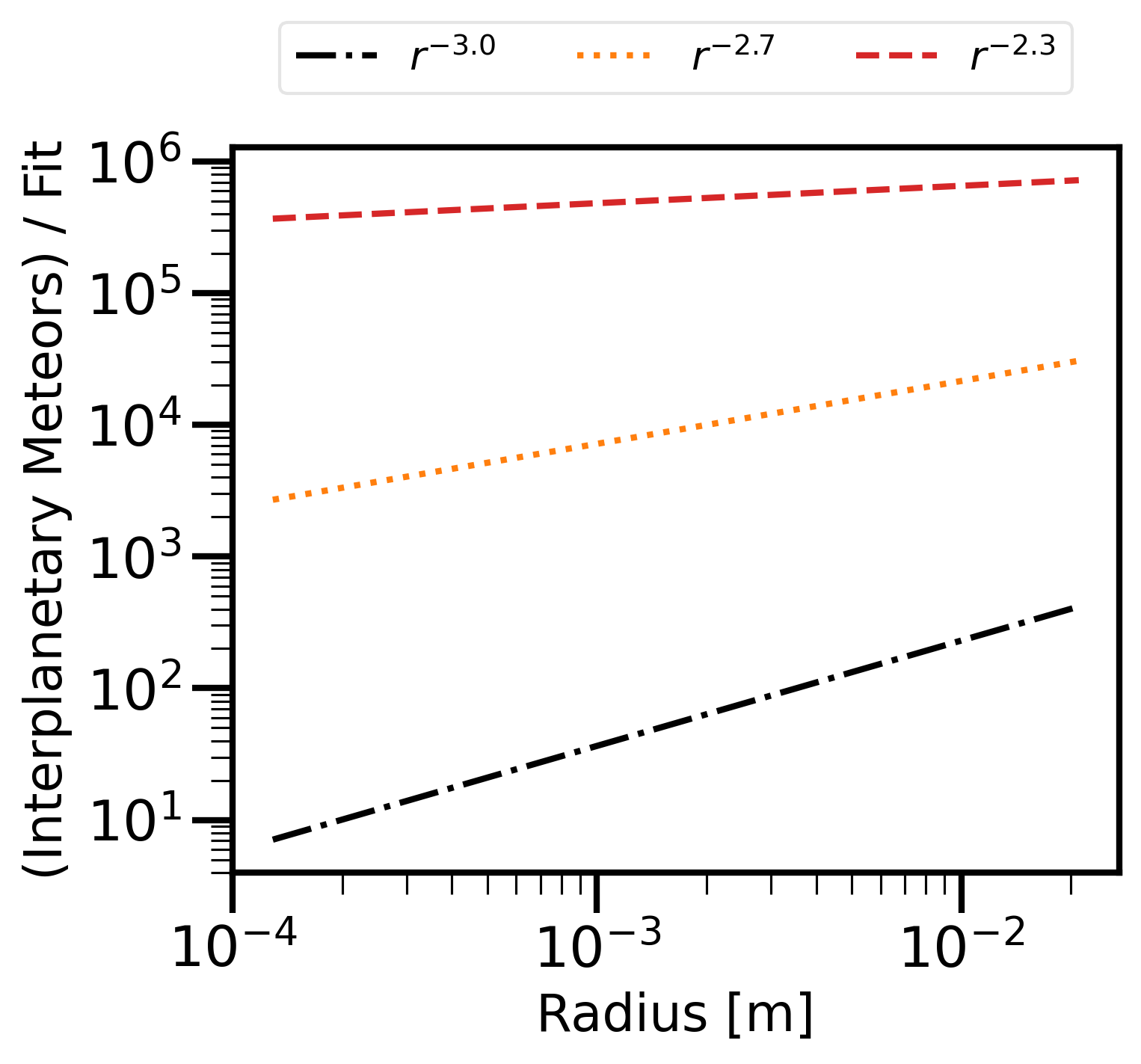}
    \caption{Interplanetary-to-interstellar ratio of the measured differential flux of observed meteors on Earth to the corresponding derived power-law fit for interstellar impactors. The linear fit fluxes are depicted for $r^{-3.0}$ \citep{Jewitt2023ARAA} to match 1I and 3I with ISD detections by spacecraft, $r^{-2.7}$ to match 1I and 3I with hyperbolic radar meteors, and $r^{-2.3}$ to match 1I and 3I with upper limit hyperbolic optical meteors (see Figure~\ref{fig:fluxes}).
    }
    \label{fig:ratio_meteors}
\end{figure}

Figure~\ref{fig:ratio_meteors} shows a comparison of the power‐law fit of \(r^{-3.0}\), as derived from dust detections of spacecraft and kilometer‐scale ISO detections, would predict roughly one interstellar meteoroid per \(10\) to \(10^{3}\) interplanetary meteors -- a ratio not supported by observations. Constraining the flux fit to the kilometer-sized ISOs with the radar‐derived slope of \(r^{-2.7}\) or with the optical upper limit of \(r^{-2.3}\) instead predicts one interstellar meteoroid per \(10^{3}\!-\!10^{4}\) or \(10^{5}\!-\!10^{6}\) interplanetary meteors, respectively. The last values are more compatible with current meteor survey observations.

This comparison assumes a single reservoir for the kilometer‐scale ISO. The cometary activity of 2I/Borisov and 3I/ATLAS, contrasted with the lack of detected volatiles in 1I/‘Oumuamua, suggests that both cometary and asteroidal components likely contribute, with potentially different fluxes. This distinction will be better constrained as ISO statistics improve. It may also prove worthwhile to compare these fluxes with those implied by exocomet size-frequency distribution estimates for $\beta$ Pictoris \citep{LecavelierdesEtangs2022} and RZ Psc \citep{Gibson2025}.

Plotting spacecraft and kilometer-sized fluxes together produces an apparent alignment on a log–log diagram \(r^{-3.0}\). However, this may be coincidental given the large uncertainties and different physical origins and thus produce unrealistic predictions of interstellar meteoroids. The flux of submicron ISD detected by spacecraft from the LIC exceeds the expected flux of similarly-sized grains ejected from exoplanetary systems by several orders of magnitude. This implies that (i) there is a dominant local interstellar component of only small particles or (ii) the spacecraft measurements need to be reassessed, especially since the Ulysses and Galileo grain size distribution appears inconsistent with that inferred from interstellar extinction \citep{Draine2009SSRv143333D}.

%%%%%%%%%%%%%%%%%%%%%%%%%%%%%%%%%%%%%%%%%%%%%%%%%%%%
\section{Conclusion}

In this paper, we have estimated the flux of ISOs implied by the detections of macroscopic ISOs and compared it with the fluxes derived from meteor surveys and spacecraft ISD measurements.
The analyses presented here reveal a gap in our understanding of the incoming interstellar population. This could be best interpreted as a consequence of mixing distinct source populations, destruction processes, and transport regimes rather than as evidence for a single, continuous size-frequency distribution. 
Future progress requires (i) large-area dust instruments with precise trajectory reconstruction to discriminate interstellar from interplanetary grains and (ii) meteor surveys with improved velocity accuracy, achieved through global coverage by higher temporal resolution instruments with long focal length optics, such as the Canadian Automated Meteor Observatory \citep[CAMO;][]{Weryk2013Icar225614W}, to validate or reject hyperbolic events.

\begin{acknowledgements}
      EP-A acknowledges support from the LUMIO project funded by the Agenzia Spaziale Italiana (2024-6-HH.0). DZS is supported by an NSF Astronomy and Astrophysics Postdoctoral Fellowship under award AST-2303553. This research award is partially funded by a generous gift of Charles Simonyi to the NSF Division of Astronomical Sciences. The award is made in recognition of significant contributions to Rubin Observatory’s Legacy Survey of Space and Time.
\end{acknowledgements}

\bibliographystyle{aa}
\bibliography{bibliography}

\begin{appendix} 
\section{Additional material}

Supplementary data that complement the discussion in the main text are presented in this section: Table~\ref{tab:interstellar} summarizes the properties of the candidate interstellar meteoroids and the confirmed ISOs considered in this study.

\begin{table*}
\centering
\caption{Properties of candidate interstellar meteoroids and confirmed ISOs. }
\small
\begin{tabular}{lrrrrrrrr}
\hline
Type & Event & $e$ & $i\ [\mathrm{deg}]$ & $q\ [\mathrm{AU}]$ & $V_{\infty}\ [\mathrm{km\,s^{-1}}]$ & $\alpha_g\ [\mathrm{deg}]$ & $\delta_g\ [\mathrm{deg}]$ & $\sigma_{\mathrm{hyper}}$ \\
\hline
Radar meteor & 2014-268-1026 & 1.12 & 16.6 & 0.26 & 21 & 100.3 & 14.3 & 3.78 \\
 &  & [0.03] & [1.2] & [0.02] & [2] & [1.4] & [0.7] &  \\
Radar meteor & 2015-008-1D0C & 1.16 & 28.2 & 0.998 & 13 & 192 & 85.2 & 3.33 \\
 &  & [0.05] & [0.8] & [0.0] & [2] & [9] & [1.0] &  \\
Radar meteor & 2014-004-0805 & 1.17 & 67 & 0.34 & 22 & 196.8 & -38 & 3.33 \\
 &  & [0.05] & [3] & [0.04] & [2] & [0.8] & [2] &  \\
Radar meteor & 2017-283-2484 & 1.12 & 22 & 0.289 & 21 & 115.4 & 9.1 & 3.51 \\
 &  & [0.04] & [2] & [0.010] & [3] & [0.6] & [1.3] &  \\
Radar meteor & 2014-299-0152 & 1.11 & 14.8 & 0.227 & 22 & 136.8 & 23.8 & 3.19 \\
 &  & [0.04] & [1.3] & [0.012] & [3] & [1.2] & [0.6] &  \\
Optical meteor & EN200117\_032343 & 1.034 & 150.58 & 0.9378 & 9.6 & 220.94 & 1.95 & 2.50 \\
 &  & [0.014] & [0.07] & [0.0009] & [0.7] & [0.07] & [0.02] &  \\
Optical meteor & EN220117\_044857 & 1.007 & 146.22 & 0.717 & 8.2 & 234.16 & -1.44 & 0.61 \\
 &  & [0.011] & [0.08] & [0.002] & [0.8] & [0.02] & [0.03] &  \\
Optical meteor & EN270117\_041236 & 1.03 & 120.8 & 0.819 & 9.7 & 208.90 & 22.75 & 1.86 \\
 &  & [0.02] & [0.2] & [0.002] & [1.0] & [0.05] & [0.06] &  \\
Optical meteor & EN280117\_011703 & 1.04 & 105.9 & 0.98410 & 10 & 231.54 & 27.6 & 1.17 \\
 &  & [0.04] & [0.4] & [0.00010] & [2] & [0.04] & [0.2] &  \\
Optical meteor & EN220717\_215951 & 1.00 & 121.3 & 0.9842 & 2 & 20.86 & 46.3 & 0.14 \\
 &  & [0.02] & [0.2] & [0.0008] & [3] & [0.12] & [0.2] &  \\
Optical meteor & EN150817\_005021 & 1.009 & 168.04 & 0.560 & 4.8 & 23.18 & 15.69 & 2.14 \\
 &  & [0.004] & [0.10] & [0.002] & [0.7] & [0.04] & [0.05] &  \\
Optical meteor & EN200917\_182805 & 1.009 & 147.92 & 0.7339 & 5.8 & 63.05 & 38.079 & 1.54 \\
 &  & [0.006] & [0.03] & [0.0011] & [0.6] & [0.02] & [0.008] &  \\
Optical meteor & EN280917\_231320 & 1.00 & 138.8 & 0.859 & 5.3 & 78.24 & 45.89 & 0.08 \\
 &  & [0.02] & [0.2] & [0.002] & [1.5] & [0.08] & [0.09] &  \\
Optical meteor & EN171017\_024843 & 1.04 & 167.38 & 0.803 & 8.7 & 99.514 & 29.996 & 2.29 \\
 &  & [0.02] & [0.03] & [0.002] & [1.0] & [0.005] & [0.007] &  \\
Optical meteor & EN011117\_003335 & 1.014 & 117.62 & 0.9317 & 7.5 & 159.78 & 47.75 & 1.21 \\
 &  & [0.011] & [0.09] & [0.0006] & [0.7] & [0.06] & [0.02] &  \\
Optical meteor & EN121117\_034640 & 1.004 & 127.60 & 0.8952 & 7.2 & 121.77 & -10.14 & 1.06 \\
 &  & [0.004] & [0.06] & [0.0005] & [0.2] & [0.03] & [0.04] &  \\
Optical meteor & EN271117\_012837 & 1.048 & 151.07 & 0.9140 & 10.0 & 159.89 & -9.36 & 5.64 \\
 &  & [0.009] & [0.14] & [0.0007] & [0.4] & [0.04] & [0.09] &  \\
Optical meteor & EN121217\_001715 & 1.012 & 59.4 & 0.3894 & 9.2 & 114.23 & 53.91 & 2.97 \\
 &  & [0.004] & [0.2] & [0.0005] & [0.5] & [0.07] & [0.02] &  \\
Optical meteor & EN120918\_222017 & 1.012 & 109.44 & 0.99880 & 5.5 & 81.51 & 65.42 & 1.06 \\
 &  & [0.012] & [0.09] & [0.00010] & [0.9] & [0.04] & [0.02] &  \\
Optical meteor & EN150918\_005203 & 1.007 & 91.00 & 0.574 & 5.5 & 19.86 & 50.05 & 1.08 \\
 &  & [0.006] & [0.14] & [0.002] & [0.8] & [0.04] & [0.04] &  \\
Optical meteor & EN190918\_213159 & 1.028 & 158.83 & 0.497 & 8.6 & 54.47 & 29.120 & 3.72 \\
 &  & [0.008] & [0.06] & [0.003] & [0.8] & [0.02] & [0.013] &  \\
Optical meteor & EN290918\_223439 & 1.018 & 116.06 & 0.727 & 7.0 & 61.91 & 54.26 & 1.62 \\
 &  & [0.011] & [0.13] & [0.002] & [1.0] & [0.04] & [0.02] &  \\
Optical meteor & EN141218\_032548 & 1.018 & 151.10 & 0.9037 & 8.7 & 169.55 & 22.36 & 1.70 \\
 &  & [0.011] & [0.05] & [0.0006] & [0.6] & [0.02] & [0.02] &  \\
Kilomete scale & 1I/‘Oumuamua & 1.20113 & 122.7417 & 0.255912 & 26.4053 & 280.50 & 34.30 & $\sim$10$^4$ \\
 &  & [0.00002] & [0.0003] & [0.000007] & [0.0010] & [0.08] & [0.08] &  \\
Kilomete scale & 2I/Borisov & 3.35648 & 44.05264 & 2.006521 & 32.27771 & 32.7894 & 59.44010 & $\sim$10$^5$ \\
 &  & [0.00002] & [0.00002] & [0.000003] & [0.00011] & [0.0002] & [0.00010] &  \\
Kilomete scale & 3I/ATLAS & 6.141 & 175.1132 & 1.3566 & 57.982 & 295.043 & -19.0704 & $\sim$10$^3$ \\
 &  & [0.002] & [0.0002] & [0.0003] & [0.005] & [0.004] & [0.0006] &  \\
\hline
\label{tab:interstellar}
\end{tabular}
\tablefoot{The five radar meteors from \citet{Froncisz2020PSS19004980F} and the optical fireballs 
from \citet{Borovicka2022AA667A158B} are shown alongside 1I/‘Oumuamua, 2I/Borisov and 
3I/ATLAS, obtained from \citet{JPLHorizonsOumuamua,JPLHorizonsBorisov,JPLHorizonsATLAS}. 
Displayed parameters are eccentricity, inclination, perihelion distance, heliocentric 
hyperbolic excess velocity (computed using the vis-viva equation), geocentric equatorial radiant, and hyperbolic significance.}
\end{table*}

\end{appendix}

\end{document}